\documentclass{llncs}

\def\Delsig{\Delta-\Sigma}
\usepackage{graphicx,color}
\usepackage{amssymb}

\begin{document}

\newcommand{\nats}{\mathbb{N}}
\newcommand{\reals}{\mathbb{R}}
\newcommand{\ints}{\mathbb{Z}}
\newcommand{\infi}{\mbox{\it inf\/}}
\def\Sys{\mathcal{S}}
\pagestyle{empty}

\title{Statistical Model Checking : An Overview}
\titlerunning{Statistical Model Checking : An Overview}
\author{Axel Legay, Beno\^it Delahaye}
\authorrunning{Legay}
\institute{INRIA/IRISA, Rennes\\
\email{alegay@irisa.fr}}

\maketitle              % typeset the title of the contribution

\begin{abstract}
  Quantitative properties of stochastic systems are usually specified
  in logics that allow one to compare the measure of executions
  satisfying certain temporal properties with thresholds. The model
  checking problem for stochastic systems with respect to such logics
  is typically solved by a numerical
  approach\,\cite{KNP04a,CG04,Var85,JKOSZ07,HWZ08,BRV04} that
  iteratively computes (or approximates) the exact measure of paths
  satisfying relevant subformulas; the algorithms themselves depend on
  the class of systems being analyzed as well as the logic used for
  specifying the properties. Another approach to solve the model
  checking problem is to \emph{simulate} the system for finitely many
  runs, and use \emph{hypothesis testing} to infer whether the samples
  provide a \emph{statistical} evidence for the satisfaction or
  violation of the specification. In this short paper, we survey the
  statistical approach, and outline its main advantages in terms of
  efficiency, uniformity, and simplicity.
\end{abstract}

\section{Introduction and Context}

Quantitative properties of stochastic systems are usually specified in
logics that allow one to compare the measure of executions satisfying
certain temporal properties with thresholds. The model checking
problem for stochastic systems with respect to such logics is
typically solved by a numerical approach that iteratively computes (or
approximates) the exact measure of paths satisfying relevant
subformulas. The algorithm for computing such measures depends on the
class of stochastic systems being considered as well as the logics
used for specifying the correctness properties. Model checking
algorithms for a variety of contexts have been
discovered~\cite{BHHK03,CY95,CG04} and there are mature tools (see
e.g.~\cite{KNP04,CB06}) that have been used to analyze a variety of
systems in practice.

Despite the great strides made by numerical model checking algorithms,
there are many challenges. Numerical algorithms work only for special
systems that have certain structural properties. Further the
algorithms require a lot of time and space, and thus scaling to large
systems is a challenge. Finally, the logics for which model checking
algorithms exist are extensions of classical temporal logics, which
are often not the most popular among engineers.

Another approach to verify quantitative properties of stochastic
systems is to \emph{simulate} the system for finitely many runs, and
use \emph{hypothesis testing} to infer whether the samples provide a
\emph{statistical} evidence for the satisfaction or violation of the
specification~\cite{YS02}. The crux of this approach is that since
sample runs of a stochastic system are drawn according to the
distribution defined by the system, they can be used to get estimates
of the probability measure on executions. Starting from time-bounded
PCTL properties~\cite{YS02}, the technique has been extended to handle
properties with unbounded until operators~\cite{SVA05a}, as well as to
black-box systems~\cite{SVA04,You05b}. Tools based on this idea have
been built~\cite{SVA05b,You05c}, and they have been used to analyze
many systems.

This approach enjoys many advantages. First, these algorithms only
require that the system be executable (or rather, sample executions be
drawn according to the measure space defined by the system). Thus, it
can be applied to larger class of systems than numerical model
checking algorithms including black-box systems and infinite state
systems. Second the approach can be generalized to a larger class of
properties, including Fourier transform based logics. Finally, the
algorithm is easily parallelizable, which can help scale to large
systems. However, the statistical approach also has some disadvantages
when compared with the numerical approach. First, it only provides
probabilistic guarantees about the correctness of the algorithms
answer. Next, the sample size grows very large if the model checker's
answer is required to be highly accurate. Finally, the statistical
approach only works for purely probabilistic systems, i.e., those that
do not have any nondeterminism. Furthermore, since statistical tests
are used to determine the correctness of a system, the approach only
works for systems that ``robustly'' satisfy a given property, i.e.,
the actual measure of paths satisfying a given subformula, is bounded
away from the thresholds to which it is compared in the specification.

In this short paper, we will overview some of existing statistical
model checking algorithms and discuss their efficiency. We will
present the hypothesis testing algorithms that are at the heart of
most of statistical algorithms and show how to uniformly analyze a
large class of systems and properties. We will also discuss case
studies.

\begin{quote}
  Most of these results are taken from Haakan Younes PhD
  Thesis\,\cite{You05a}.
\end{quote}

\section{What do we want to do?}

We consider a stochastic system $\Sys$ and a property $\phi$. An {\em
  execution} of $\Sys$ is a possibly infinite sequence of states of
$\Sys$. Our objective is to solve the {\em probabilistic model
  checking problem}, i.e., to decide whether $\Sys$ satisfies $\phi$
with a probability greater or equal to a certain threshold
$\theta$. The latter is denoted $S\models P_{{\geq}\theta}(\phi)$,
where $P$ is called a {\em probabilistic operator}. This paper will
overview solutions to this problem. These solutions depend on the
nature of $\Sys$ and $\phi$. We consider three cases.

\begin{enumerate}
\item We first assume that $\Sys$ is a {\em white-box system}, i.e.,
  that one can generate as much executions of the system as we
  want. We also assume that $\phi$ does not contain probabilistic
  operators. In Section \ref{sec:begin}, we recall basic statistical
  algorithms that can be used to verify bounded properties (i.e.,
  properties that can be verified on fixed-length execution) of
  white-box systems.
\item In Section \ref{sec:logic}, we discuss extensions to the full
  probabilistic computation tree logic\cite{CG04}. There, we consider the
  case where $\phi$ can also contain probabilistic operators and the
  case where it has to be verified on infinite executions.
\item In Section \ref{sec:black}, we briefly discuss the verification
  of {\em black-box systems}, i.e. systems for which a part of the
  probability distribution is unknown.
\end{enumerate}

\noindent
In addition, in Section \ref{expe}, we will present various experiments
that show that (1) statistical model checking algorithms are more
efficient than numerical ones, and (2) statistical model checking
algorithms can be applied to solve problems that are beyond the scope
of numerical methods. Finally, Section \ref{sec:future} discusses the future of
statistical model checking.

\begin{remark}
  The objective of the present tutorial is not to feed the reader with
  technical details, but rather to present the concepts of statistical
  model checking, and outline its main advantages in terms of
  efficiency, uniformity, and simplicity.
\end{remark}

\begin{remark}
  There are other techniques that allow to estimate the probability
  for $\Sys$ to satisfies $\phi$. Those techniques, which are based on
  Monte-Carlo techniques, will not be presented in this paper. The
  interested reader is redirected to \cite{GS05,HLMP04,LMPR07} for
  more details.
\end{remark}

\section{Statistical Model Checking : The Beginning}
\label{sec:begin}

In this section, we overview several statistical model checking
techniques. We assume that $\Sys$ is a white-box system and that
$\phi$ is a bounded property. By bounded properties, we mean
properties that can be defined on finite executions of the system. In
general, the length of such executions has to be pre-computed.

Let $B_i$ be a discrete random variable with a Bernoulli distribution
of parameter $p$. Such a variable can only take $2$ values $0$ and $1$
with $Pr[B_i=1]=p$ and $Pr[B_i=0]=1-p$. In our context, each variable
$B_i$ is associated with one simulation of the system. The outcome for
$B_i$, denoted $b_i$, is $1$ if the simulation satisfies $\phi$ and
$0$ otherwise. To make sure that the above approach works, one has to
make sure that one can get the result of any experiment in a finite
amount of time. In general, this means that we are considering bounded
properties, i.e., properties that can be decided on finite executions.

\begin{remark}
  All the results presented in this section are well-known
  mathematical results coming from the area of statistics. As we shall
  see, these results are sufficient to verify bounded properties of a
  large class of systems. As those properties are enough in many
  practical applications, one could wonder whether the contribution of
  the computer scientist should not be at the practical level rather
  than at the theoretical one.
\end{remark}

% \begin{remark}
%   The reader shall observe that the following approach offers a
%   uniform solution to the probabilistic model checking problem of
%   bounded properties. With the numerical approaches, one would end up
%   with an algorithm per model.
% \end{remark}

Before going further one should answer one last question: {\em ``What
  is the class of models that can be considered?''} In fact, the
answer is quite simple: any stochastic system on which one can define
a {\em probability space} for the property under consideration. Hence,
statistical model checking provides a uniform approach for the
verification of a wide range of stochastic models, including Markov
Chains or Continuous Timed Markov Chains. In general, one does not
make the hypothesis that the system has the Markovian
property{\footnote{i.e., that the probability to go to one state only
    depends on the state in where we are, not on the history of the
    execution.}, except when working with nested formulas (see Section
  \ref{sec:logic}). There is a big warning: the technique cannot be
  used to verify properties of models that combine both
  nondeterministic and stochastic aspects. Indeed, the
  simulation-based approach could not distinguish between the
  probability distributions that are sampled.

\subsection{Qualitative Answer using Statistical Model Checking}

% The idea is to decide
% between the two hypothesis by conducting a sample set of simulation
% ons and then decide whether the
% system satisfies $Pr_{\geq \theta}(\phi)$ based on the number of
% simulations for which $\phi$ holds compared to the total number of
% simulations in the sample set.

The main approaches\,\cite{You05a,SVA04} proposed to answer the
qualitative question are based on {\em hypothesis testing}. Let
$p=Pr(\phi)$, to determine whether $p \geq \theta$, we can test $H:p
\geq \theta$ against $K:p < \theta$. A test-based solution does not
guarantee a correct result but it is possible to bound the probability
of making an error. The {\em strength} $(\alpha,\beta)$ of a test is
determined by two parameters, $\alpha$ and $\beta$, such that the
probability of accepting $K$ (respectively, $H$) when $H$
(respectively, $K$) holds, called a Type-I error (respectively, a
Type-II error ) is less or equal to $\alpha$ (respectively,
$\beta$).

A test has {\em ideal performance} if the probability of the Type-I
error (respectively, Type-II error) is exactly $\alpha$ (respectively,
$\beta$). However, these requirements make it impossible to ensure a
low probability for both types of errors simultaneously (see
\cite{You05a} for details). A solution to this problem is to relax the
test by working with an {\em indifference region} $(p_1,p_0)$ with
$p_0{\geq}p_1$ ($p_0-p_1$ is the {\em size of the region}). In this
context, we test the hypothesis $H_0: p\, {\geq} \,p_0$ against $H_1:p
\, {\leq} \,p_1$ instead of $H$ against $K$. If the value of $p$ is
between $p_1$ and $p_0$ (the indifference region), then we say that
the probability is sufficiently close to $\theta$ so that we are
indifferent with respect to which of the two hypotheses $K$ or $H$ is
accepted. The thresholds $p_0$ and $p_1$ are generally defined in term
of the single threshold $\theta$, e.g., $p_1=\theta-\delta$ and
$p_0=\theta+\delta$. We now need to provide a test procedure that
satisfies the requirements above. In the next two subsections, we
recall two solutions proposed by Younes in \cite{You05a,YS06}. 

\paragraph{Single Sampling Plan.}  

To test $H_0$ against $H_1$, we specify a constant $c$. If
$\sum_{i=1}^n b_i$ is larger than $c$, then $H_0$ is accepted, else
$H_1$ is accepted. The difficult part in this approach is to find
values for the pair $(n,c)$, called a {\em single sampling plan (SSP
  in short)}, such that the two error bounds $\alpha$ and $\beta$ are
respected. In practice, one tries to work with the smallest value of
$n$ possible so as to minimize the number of simulations
performed. Clearly, this number has to be greater if $\alpha$ and
$\beta$ are smaller but also if the size of the indifference region is
smaller. This results in an optimization problem, which generally does
not have a closed-form solution except for a few special
cases\,\cite{You05a}. In his thesis\,\cite{You05a}, Younes proposes a
binary search based algorithm that, given $p_0,p_1,\alpha, \beta$,
computes an approximation of the minimal value for $c$ and $n$.

\begin{remark}
  There are many variants of this algorithm. As an example, in
  \cite{SVA05a}, Sen et al. proposes to accept $H_0$ if
  $\frac{(\sum_{i=1}^n b_i)}{n}{\geq}p$. Here, the difficulty is to
  find a value for $n$ such that the error bounds are valid.
\end{remark}

\paragraph{Sequential probability ratio test.} The sample size for a
single sampling plan is fixed in advance and independent of the
observations that are made. However, taking those observations into
account can increase the performance of the test. As an example, if we
use a single plan $(n,c)$ and the $m>c$ first simulations satisfy the
property, then we could (depending on the error bounds) accept $H_0$
without observing the $n-m$ other simulations. To overcome this
problem, one can use the {\em sequential probability ratio test (SPRT
  in short)} proposed by Wald\,\cite{Wal45}. The approach is briefly
described below.

In SPRT, one has to choose two values $A$ and $B$ ($A>B$) (see bellow)
that ensure that the strength of the test is respected.  Let $m$ be
the number of observations that have been made so far. The test is
based on the following quotient:

\begin{equation}
\label{eq:test}
\frac{p_{1m}}{p_{0m}}=\prod_{i=1}^{m}\frac{Pr(B_i=b_i\mid
  p=p_1)}{Pr(B_i=b_i\mid
  p=p_0)}=\frac{p_1^{d_m}(1-p_1)^{m-d_m}}{p_0^{d_m}(1-p_0)^{m-d_m}},
\end{equation}
\noindent
where $d_m = \sum_{i=1}^m b_i$. The idea behind the test is to accept
$H_0$ if $\frac{p_{1m}}{p_{0m}} \geq A$, and $H_1$ if
$\frac{p_{1m}}{p_{0m}} \leq B$. The SPRT algorithm computes
$\frac{p_{1m}}{p_{0m}}$ for successive values of $m$ until either
$H_0$ or $H_1$ is satisfied; the algorithm terminates with probability
$1$\cite{Wal45}. This has the advantage of minimizing the number of
simulations. In his thesis\,\cite{You05a}, Younes proposed a
logarithmic based algorithm SPRT that given $p_0,p_1,\alpha$ and
$\beta$ implements the sequential ratio testing procedure.\\
\newline
{\bf Discussion.} Computing ideal values $A_{id}$ and $B_{id}$ for $A$
and $B$ in order to make sure that we are working with a test of
strength $(\alpha,\beta)$ is a laborious procedure (see Section 3.4 of
\cite{Wal45}). In his seminal paper\,\cite{Wal45}, Wald showed that if
one defines $A_{id}{\geq}A=\frac{(1-\beta)}{\alpha}$ and $B_{id}\leq
B=\frac{\beta}{(1-\alpha)}$, then we obtain a new test whose strength
is $(\alpha',\beta')$, but such that $\alpha'+\beta' \leq
\alpha+\beta$, meaning that either $\alpha'{\leq}\alpha$ or $\beta'
\leq \beta$. In practice, we often find that both inequalities
hold. This is illustrated with the following example taken from
\cite{You05a}.

\begin{example}
  Let $p_0=0.5$, $p_1=0.3$, $\alpha=0.2$ and $\beta=0.1$. If we use
  $A_{id}{\geq}A=\frac{(1-\beta)}{\alpha}$ and $B_{id}\leq
  B=\frac{\beta}{(1-\alpha)}$, then we are guaranteed that
  $\alpha'{\leq}0.222$ and $\beta'{\leq}0.125$. Through computer
  simulation (repruding the same experiments $10 000$ of time), we
  observe that $\alpha'{\leq}0.175$ and $\beta'{\leq}0.082$. So the
  strength of the test is in reality better than the theoretical
  assumption.
\end{example}

\subsection{Some Generalities Regarding Efficiency}

The efficiency of the above algorithms is characterized by the number
of simulations needed to obtain an answer as well as the time it costs
to compute a simulation. The latter often depends on the property
under verification.  Both numbers are {\em expected numbers} as they
change from executions to executions and can only be estimated (see
\cite{You05a} for an explanation). However, some generalities are
known. For example, it is known that, except for some situations, SPRT
is always faster than SSP. When $\theta=1$ (resp. $\theta=0$) SPRT
degenerates to SSP; it is not a problem since SSP is known to be
optimal for such values. Observe that the time complexity of
statistical model checking is independent from the state-space and
that the space complexity is of the order of the state space. Also,
the expected number of simulations for SSP is logarithmic with respect
to $\alpha$ and $\beta$ and linear with respect to the indifference
region; for SPRT, the number depends on the probability distribution
$p$.\\
\newline
An interesting discussion on complexity of statistical model
checking can be found in Section $5.4$ of \cite{You05a}.

\section{Statistical Model Checking: The Computer Science Contribution}
\label{sec:logic}
In the previous section, we have proposed statistical model checking
algorithms for verifying bounded properties of white-box systems. In
this section, we go one step further and consider three nontrivial
extensions that are:

\begin{enumerate}
\item The nested case, i.e., the case where $\phi$ can also contain
  probabilitistic operators. Example: $P_{{\geq}\theta_1}(q\Rightarrow
  P_{{\geq}\theta_2}(\phi_2))$
\item The unbounded case, i.e., the case where $\phi$ cannot be decide
  on a finite execution. Here we will restrict ourselves to the until
  property. Given two formulas $\phi_1$ and $\phi_2$, the until
  operator ensures that $\phi_1$ is true until $\phi_2$ has been seen
  (and this must happen!).
\item Boolean combinations of formulae, i.e., formulae of the form:
  $P_{{\geq}\theta_1}(\phi_1)\wedge P_{{\geq}\theta_2}(\phi_2)$.
\end{enumerate}

\noindent
We will only survey these results and give pointers to relevant
papers.

\subsection{The Unbounded Case: Until}

We are now concerned with the verification of the until property. The
property requires that a property $\phi_1$ remains valid until a
property $\phi_2$ has been seen. The problem is that we do not know a
priori the moment when $\phi_2$ will be satisfied. Hence, one has to
reason on infinite execution.  There are two works on this topics, one
by Sen et al.\cite{SVA05a} and one more recent work by Pekergin et
al. \,\cite{RP09}. We will not give details on these works, but the
reader should know that Sen works by extending the model with extra
probabilities, which makes the solution extremely slow. Pekergin uses
the new technique of {\em perfect simulation}, which is (according to
her experiments) not only faster than Sen's one, but also more general
as it allows to study the steady-state operator for continuous timed
Markov Chains.

\begin{remark}
  Contrary to the numerical results \,\cite{Var85,BRV04} The above
  results are not sufficient to verify properties of the form
  $P_{{\geq}\theta}(\phi)$, where $\phi$ is a property expressed in
  Linear Temporal Logic\,\cite{Pnu77}. Incomplete results regarding
  the verification of these properties with simulation-based
  techniques can be found in \cite{HLMP04,GS05}.
\end{remark}

\subsection{Nested Probability Operators}

We consider the problem of checking whether $\Sys$ satisfies $\phi$
with a probability greater or equal to $\theta$. However, contrary to
what we have been doing so far, we will now assume that $\phi$ cannot
be decided on a single execution, i.e., we will assume that $\phi$ is
of the form $P_{{\geq}\theta_1}\phi_1$. So, where is the difficulty?
The difficulty is that $\phi$ cannot be model checked on a single
execution, but rather depends on another test. Hence, we have to
provide a way to nest tests. In his thesis, Younes proposed the
following theorem.

\begin{theorem}
  Let $\psi=P_{{\geq}\theta}(\phi)$ be a property and assume that
  $\phi$ can be verified with Type-I error $\alpha'$ and Type-II error
  $\beta'$, then $\psi$ can be verified with Type-I error $\alpha$ and
  Type-II error $\beta$, assuming that the indifference region is of
  size at least
  $((\theta+\delta)(1-\alpha'),(1-(1-(\theta-\delta)))(1-\beta')$.
\end{theorem}

Hence one has to find a compromise between the size of the
indifference region of the inner test and the outer one. There are two
interesting facts to know about nested operators:

\begin{enumerate}
\item Even for bounded properties, the above result (and in fact, any
  result in the literature\,\cite{SVA05a,You05a,You05b,You05c}) only
  works for systems that have the Markovian property.
\item In practice, the complexity (in term of number of sampling)
  becomes exponential in the number of tests.
\end{enumerate}

\begin{remark}
  An interesting research direction would be to study the link with
  probabilistic testing\,\cite{LS91}.
\end{remark}

\subsection{Boolean Combinations}

We have to consider two operations, namely conjunction and negation
(as it is known that any Boolean combination reduces to combinations
of these two operators). We recall some results provided by Younes. We
start with conjunction.

\begin{theorem}
  Let $\psi$ be the conjunction of $n$ properties
  $\phi_1,\dots,\phi_2$. Assume that each $\phi_i$ can be decided with
  Type-I error $\alpha_i$ and Type-II error $\beta_I$. Then $\phi$ can
  be decided with Type-I error min$_i$($\alpha_i$) and Type-II error
  max$_i$($\beta_i)$.
\end{theorem}

\noindent
The idea behind the proof of the theorem is that
\begin{enumerate}
\item 
  If we claim that the conjunction is not satisfied, this means that
  we have deduced that one of the operands is not.
\item 
  If we claim that the conjunction is satisfied, this means that we
  have concluded that all the operands are satisfied. As we may have
  made mistakes in each individual verification, we get max$_i$($\beta_i)$.
\end{enumerate}

For negation, the result is provided by the following theorem.

\begin{theorem}
  To verify a formula $\neg\psi$ with Type-I error $\alpha$ and
  Type-II error $\beta$, it is sufficient to verify $\psi$ with Type-I
  error $\beta$ and Type-II error $\alpha$.
\end{theorem}

\section{Black-box Systems: a note}
\label{sec:black}

Black-box Systems is an interesting class of stochastic systems whose
treatment is beyond the scope of numerical techniques. Roughly
speaking, a black-box systems is simply a system whose probability
distribution (i.e., set of behaviors) is not totally known and cannot
be observed. Hence, one can view a black-box system as a finite set of
executions pre-computed and for which no information is available.

In the context of such systems, Type errors and indifference region
cannot play a role. Indeed, those parameters influence the number of
simulations that can be computed, but here the simulations are given
and you cannot compute more!

A solution to this problem is to conduct a SSP test, without
indifference region (i.e., $\delta$ set to $0$) and assuming that the
parameter $n$ is fixed to the number of simulations that are given in
advance. The difficulty is to chose the constant $c$ in such a way
that it becomes roughly equal to accept $H_0$ or $H_1$ if
$\theta=p$. In his thesis\,\cite{You05a} and in \cite{You06}, Younes
proposed a solution to the problem. He also shown that a previous
solution proposed by Sen \,\cite{SVA04} is not correct.

There are techniques to verify nested formulas over black-box
systems. There exists no technique for the verification of unbounded
properties. Hence there is still a lot of research to conduct in this
area.

\section{Tools and Experiments}
% 1 hour
\label{expe}

At the origin, there are two tools that implements statistical model
checking algorithms, namely {\em Ymer}\cite{You05c} and {\em
  Vesta}\cite{SVA05b}. {\em Vesta} implements a variation of the
single sampling plan algorithm. The choice of implementing the SSP
algorithm is motivated by the fact that it is easier to parallelize as
the number of simulations to perform is known in advance. However, in
his thesis, Younes showed that sequential algorithms are also easily
parallelizable. {\em Ymer} is limited to bounded properties while {\em
  Vesta} also incorporate the unbounded until. In \cite{JKOSZ07}, the
authors conducted several experiments that tend to show that (1) {\em
  Ymer} is faster than {\em Vesta} and (2) {\em Vesta} makes more
false positive (selecting the bad hypothesis) than {\em
  Ymer}. Regarding the unbounded case, it seems that {\em Vesta} is
not very efficient and can make a lot of false positive. Both {\em
  Vesta} and {\em Ymer} have been applied to huge case studies. A
comparison of {\em Ymer} and {\em Vesta} with established tools such
{\em PRISM}\,\cite{KNP04} can be found in \cite{JKOSZ07}.  \newline
There are a wide range of situations for which the bounded case
suffices. We have written a series of recent papers in where we
propose applications of SSP and SPRT to interesting problems. In the
rest of this section, we briefly recap the content of these papers.

\subsection{Verifying Circuits}

In \cite{CDL08,CDL09}, we applied SPRT to verifying properties of {\em
  mixed-signal circuits}, i.e., circuits for which there is an
interaction between analog (continuous) and digital (discrete)
values. Our first contribution was to propose a version of stochastic
discrete-time event systems that fits into the framework introduced by
Younes with the additional advantage that it explicitly handles analog
and digital signals. We also introduced {\em probabilistic signal
  linear temporal logic}, a logic adapted to the specification of
properties for mixed-signal circuits in the {\em temporal} domain and
in the {\em frequency} domain. Our second contribution was the
analysis of a $\Delsig$ modulator. A $\Delsig$ modulator is an
efficient {\em Analog-to-Digital Converter circuit}, i.e., a device
that converts analog signals into digital signals. A common critical
issue in this domain is the analysis of the \emph{stability} of the
internal state variables of the circuit. The concern is that the
values that are stored by these variables can grow out of control
until reaching a maximum value, at which point we say that the circuit
\emph{saturates}. Saturation is commonly assumed to compromise the
quality of the analog-to-digital conversion. In
\cite{DangDonzeMaler04} and \cite{GuptaKR04} reachability techniques
developed in the area of hybrid systems are used to analyze the
stability of a third-order modulator. Their idea is to use such
techniques to guarantee that for \emph{every} input signal in a given
range, the states of the system remain stable. While this
reachability-based approach is sound, it has important drawbacks such
as (1) signals with long duration cannot be practically analyzed, and
(2) properties that are commonly specified in the frequency domain
rather than in the time domain cannot be checked. Our results show
that a simulation-based approach makes it possible to handle
properties and signals that are beyond the scope of the
reachability-based approach. As an example, in our experiments, we
analyze discrete-time signals with $24000$ sampling points in seconds,
while the approach in \cite{DangDonzeMaler04} takes hours to analyze
signals with up to $31$ sampling points. We are also able to provide
insight into a question left open in \cite{DangDonzeMaler04} by
observing that saturation does not always imply an improper signal
conversion. This can be done by comparing the Fourier transform of
each of the input analog signals with the Fourier transform of its
corresponding digital signal. Such a property can easily be expressed
in our logic and Model Checked with our simulation-based approach. We
are unaware of other formal verification techniques that can solve
this problem. Indeed, numerical techniques cannot reason on an
execution at a time.

\subsection{Systems Biology}

In \cite{CFLHJL08}, we considered the verification of complex
biological systems. we introduced a new tool, called {\sc BioLab}, for
\emph{formally} reasoning about the behavior of stochastic dynamic
models by integrating SPRT into the {\sc BioNetGen} \cite{BNG1,BNGL}
framework for rule-based modeling. We then used {\sc BioLab} to verify
the stochastic bistability of T-cell signalling. There are three more
challenges in the systems biology area (the reader is invited to think
about these problems and to check the existing literature):

\begin{enumerate}
\item
How to perform efficient simulations? 
\item
How to take into account prior knowledge on the model?
\item 
  What are the logics dedicated to biologists than can be model
  checked with the statistical approach?
\end{enumerate}

\begin{remark}
In fact, statistical model checking techniques recently received a lot
of attention in the area of systems biology. As an example, in $2009$,
Carnegie Mellon University was awarded a $10 000 000$ grant for
applying such techniques in the medical area. 
\end{remark}

\subsection{Heterogeneous applications}

In \cite{BBBCDL10}, we have proposed to apply statistical model
checking techniques to the verification of {\em heterogeneous
  applications}.  Systems integrating multiple heterogeneous
distributed applications communicating over a shared network are
typical in various sensitive domains such as aeronautic or automotive
embedded systems. Verifying the correctness of a particular
application inside such a system is known to be a challenging task,
which is often beyond the scope of existing exhaustive validation
techniques.

In our paper, we proposed to exploit the structure of the system in
order to increase the efficiency of the verification process. The idea
is conceptually simple: instead of performing an analysis of the
entire system, we proposed to analyze each application separately, but
under some particular context/execution environment.  This context is
a {\em stochastic abstraction} that represents the interactions with
other applications running within the system and sharing the
computation and communication resources.  The idea is to build such a
context automatically by simulating the system and learning the
probability distributions of key characteristics impacting the
functionality of the given application. The abstraction can easily be
analyzed with statistical model checking techniques.

The overall contribution of our study is an application of the above
method on an industrial case study, the {\em heterogeneous
  communication system} (HCS for short) deployed for cabin
communication in a civil airplane. HCS is an heterogeneous system
providing entertainment services (ex : audio/video on passengers
demand) as well as administrative services (ex: cabin illumination,
control, audio announcements), which are implemented as distributed
applications running in parallel, across various devices within the
plane and communicating through a common Ethernet-based network.  The
HCS system has to guarantee stringent requirements, such as reliable
data transmission, fault tolerance, timing and synchronization
constraints. An important requirement is the {\em accuracy of clock
  synchronization} between different devices. This latter property
states that the difference between the clocks of any two devices
should be bounded by a small constant, which is provided by the user
and depends on his needs (for example, to guarantee the fiability of
another service). Hence, one must be capable to compute the smallest
bound for which synchronization occurs and compare it with the bound
expected by the user. Unfortunately, due to the large number of
heterogeneous components that constitute the system, deriving such a
bound manually from the textual specification is an unfeasible
task. In this paper, we propose a formal approach that consists in
building a formal model of the HCS, then we apply simulation-based
algorithms to this model in order to deduce the smallest value of the
bound for which synchronization occurs. We start with a fixed value of
the bound and check whether synchronization occurs. If yes, then we
make sure that this is the best one. If no, we restart the experiment
with a new value.

We have been able to derive precise bounds that guarantee proper
synchronization for all the devices of the system. We also computed
the probability to satisfy the property for smaller values of the
bound, i.e., bounds that do not satisfy the synchronization property
with probability $1$. Being able to provide such an information is of
clear importance, especially when the best bound is too high with
respect to user's requirements. We have observed that the values we
obtained strongly depend on the position of the device in the
network. We also estimated the average and worst proportion of
failures per simulation for bounds that are smaller than the one that
guarantees synchronization. Checking this latter property has been
made easy because statistical model checking allows us to reason on
one execution at a time. Finally, we have also considered the
influence of clock drift on the synchronisation results. The
experiments highlight the generality of our technique, which could be
applied to other versions of the HCS as well as to other heterogeneous
applications.

\section{The Future of Statistical Model Checking}
\label{sec:future}
There are various directions for future research in the statistical
model checking area. Here is a list of possible topics.

\begin{itemize}
\item Using efficient techniques for performing simulation is crucial
  to guarantee good performances for any statistical model checking
  algorithm. Unfortunately, the existing algorithms do not exploit
  efficient simulation techniques. It would thus be worth combining
  statistical model checking algorithms with such techniques (example
  : rare-event simulations, , ...). This is a huge implementation
  effort which also requires to define a methodology to select the
  good simulation technique to be applied.
\item
Statistical model checking algorithms have not yet been applied to the
verification of multi-core systems, this area should be
investigated.
\item
Statistical model checking algorithms do not apply to systems that
combine both stochastic and non deterministic aspects. Extending the
results to such systems is however crucial to perform verification of
security protocols, networking protocols, and performance protocols.
\item
Statistical model checking algorithms reduce to decide between two
hypothesis. In many areas, especially systems biology, we may have a
prior knowledge on the probability to satisfy each
hypothesis. Incorporating this prior knowledge in the verification
process may considerably reduce the number of simulations needed for
the algorithm to terminate.
\item
Statistical model checking algorithms suppose that the property $\phi$
can be checked on finite executions of the system. There are however
many situations where $\phi$ cannot be checked in a finite amount of
time. This is for example the case when $\phi$ is a long-run average
or a steady state property. In systems biology, we are clearly
interested in the study of such properties.
\item Verifying applications running within a huge heterogeneous
  system without is a challenging problem. In a recent
  work\,\cite{BBBCDL10}, the authors have proposed a new
  simulation-based technique for solving such problem. The technique
  starts by performing simulations of the system in order to learn the
  context in where the application is used. Then, it creates a
  stochastic abstraction for the application, which takes the context
  information into account. Up to know, there is no automatic way to
  learn the context and derive the stochastic context. However, what
  we have observed so far is that it often takes the form of
  properties that cannot be expressed in classical temporal
  logic. Hence, statistical model checking may be our last resort to
  analyze the resulting abstraction.
\item
Statistical model checking may help testers. In \cite{MHNC09}, Cavalli
et al. proposed to use statistical techniques for conformance testing
of timed stochastic systems. The technique should be automated. This
could lead to new algorithms for verifying the so-called black-box
systems.
\end{itemize}

\section*{Acknowledgments}

We would like to thanks our collaborators on the statistical model
checking project: Sumit Jha, Marius Bozga, Saddeck Bensalem.

\bibliography{main,thesis} \bibliographystyle{abbrv}
\end{document}